# Suppression of vortical turbulence in plasma of mass-separator in crossed electrical and magnetic fields due to finite lifetime of electrons and ions and due to finite system length

*V.I.Maslov, I.P.Levchuk, I.N.Onishchenko, V.B.Yuferov*
*NSC Kharkov Institute of Physics & Technology, 61108 Kharkov, Ukraine*
*vmaslov@kipt.kharkov.ua*

The dispersion equation, describing the instability development of vortex turbulence excitation in cylindrical plasma in crossed radial electric and axial magnetic fields with taking into account the longitudinal inhomogeneity and finite time of leaving of plasma electrons and ions from the system, has been derived. It is shown that the finite length of system time and finite time of system leaving of plasma electrons and ions leads to the appearance of the instability threshold and to decrease of growth rate of its development.

## 1. INTRODUCTION

It is well known from numerous numerical simulations (see, for example, [1]) and from experiments (see, for example [2]) that electron density nonuniformity in kind of discrete vortices are long-living structures. In experiments [2] a rapid re-organization of discrete electron density nonuniformity has been observed in the spatial distribution of vorticity in pure electron plasma when a discrete vortex has been immersed in an extended distribution of the background vorticity. In plasma lens [3-11] for high-current ion beam focusing, in mass-separator [12-16], in nuclear fusion [17-19] and in plasma-optical device for the elimination of droplets in cathodic ARC plasma coating [20], a vortical turbulence has been excited in crossed radial electrical and longitudinal magnetic fields [21-23] by gradient of external magnetic field or by gradient of plasma density. This turbulence is a distributed vorticity. In this paper the excitation and damping of similar vortical turbulence, excited in cylindrical plasma in crossed radial electrical $E_{0r}$ and longitudinal magnetic $H_0$ fields, is investigated theoretically. From general nonlinear equation for vorticity the dispersion relation, which describes the instability development of vortex turbulence excitation, has been derived. It is shown that the finite length of system time and finite time of system leaving of plasma electrons and ions leads to the appearance of the instability threshold and to decrease of growth rate of its development.

## 2. EXCITATION OF VORTICES

Let us derive the dispersion relation. We take into account that the ions pass with velocity $V_{bi}$ through system of length L during time, approximately equal $\tau_i = \frac{L}{V_{bi}}$. We also take into account that the electrons pass through system and are renovated in system also during finite time, $\tau_e$. Damping of perturbations of densities and velocities of electrons and ions at recovery of their unperturbed values we describe, using $\nu_i \equiv \frac{1}{\tau_i}$, $\nu_e \equiv \frac{1}{\tau_e}$.

We use the electron hydrodynamic equations

$$\frac{\partial \vec{V}}{\partial t} + \nu_e \left( \vec{V} - \vec{V}_{\theta o} \right) + \left( \vec{V} \vec{\nabla} \right) \vec{V} =$$
$$= \left( \frac{e}{m_e} \right) \vec{\nabla} \phi + \left[ \vec{\omega}_{He}, \vec{V} \right] - \left( \frac{V_{th}^2}{n_e} \right) \vec{\nabla} n_e ,$$
$$\frac{\partial n_e}{\partial t} + \frac{(n_e - n_{oe})}{\tau_e} + \vec{\nabla} \left( n_e \vec{V} \right) = 0 . \quad (1)$$

Also we use the ion hydrodynamic equations

$$\frac{\partial \vec{V}_i}{\partial t} + \nu_i \left( \vec{V} - \vec{V}_{bi} \right) + \left( \vec{V}_i \vec{\nabla} \right) \vec{V}_i = -\left( \frac{q_i}{m_i} \right) \vec{\nabla} \phi ,$$
$$\frac{\partial n_i}{\partial t} + \frac{(n_i - n_{oi})}{\tau_i} + \nabla \left( n_i \vec{V}_i \right) = 0 . \quad (2)$$

and Poisson equation for the electrical potential, $\phi$,

$$\Delta \phi = 4\pi \left( e n_e - q_i n_i \right) . \quad (3)$$

Here $\vec{V}$, $n_e$ are the velocity and density of electrons; $V_{th}$ is the electron thermal velocity; $V_{\theta o}$ is the electron azimuth drift velocity in crossed fields; $\vec{V}_i$, $n_i$, $q_i$, $m_i$ are the velocity, density, charge and mass of ions.

As it will be visible from the further, the dimensions of the vortical perturbations are much larger than the electron Debye radius, $r_{de} \equiv \frac{V_{th}}{\omega_{pe}}$, then the last term in (1) can be neglected. Here $\omega_{pe} \equiv \left( \frac{4\pi n_{oe} e^2}{m_e} \right)^{1/2}$, $n_{oe}$ is the unperturbed electron density.

From equations (1) one can derive non-linear equations

$$d_t \left[ \frac{(\alpha - \omega_{He})}{n_e} \right] = \left[ \frac{(\alpha - \omega_{He})}{n_e} \right] \partial_z V_z - \frac{\alpha \nu_e}{n_e} , \quad (4)$$

$$d_t V_z + \nu_e V_z = \left( \frac{e}{m_e} \right) \partial_z \phi$$

describing both transversal and longitudinal electron dynamics. Here

$$d_t \equiv \partial_t + \left( \vec{V}_\perp \vec{\nabla}_\perp \right) , \quad \partial_z \equiv \frac{\partial}{\partial z} , \quad \partial_t \equiv \frac{\partial}{\partial t} , \quad (5)$$

$\vec{V}_\perp$, $V_z$ are the transversal and longitudinal electron velocities, $\alpha$ is the vorticity, the characteristic of electron vortical motion, $\alpha \equiv \vec{e}_z \text{rot} \vec{V}$.

Taking into account higher linear terms, from (1) one can obtain

$$\vec{V}_\perp \approx \left(\frac{e}{m_e \omega_{He}}\right)\left[\vec{e}_z, \vec{\nabla}_\perp \phi\right] + \left(\frac{e}{m_e \omega_{He}^2}\right)(\partial_t + \nu_e)\vec{\nabla}_\perp \phi \quad (6)$$

From (6) we derive

$$\alpha \equiv \mu \omega_{He} \approx -\frac{2eE_{or}}{rm_e \omega_{He}} - \frac{eE_{or}}{m_e}\partial_r\left(\frac{1}{\omega_{He}}\right) + \frac{e}{m_e \omega_{He}}\Delta_\perp \varphi +$$
$$+ \frac{e}{m_e}(\partial_r \varphi)\partial_r\left(\frac{1}{\omega_{He}}\right) + \frac{e}{m_e}(\partial_t + \nu_e)\vec{e}_z\left[\vec{\nabla}_\perp, \left(\frac{1}{\omega_{He}^2}\right)\vec{\nabla}_\perp \varphi\right],$$
$$\vec{\nabla}\phi \equiv \vec{\nabla}\varphi - \vec{E}_{or} \quad (7)$$

Here $E_{or}$ is the radial focusing electric field, $\varphi$ is the electric potential of the vortical perturbation;

$$-\frac{2eE_{or}}{rm_e \omega_{He}} = \left(\frac{\omega_{pe}^2}{\omega_{He}}\right)\left(\frac{\Delta n}{n_{oe}}\right) \equiv \eta \omega_{He}, \quad \Delta n \equiv n_{oe} - n_{oi}\frac{q_i}{e}.$$

From (7) $\alpha \approx \left(\frac{\omega_{pe}^2}{\omega_{He}}\right)\left(\frac{\delta n_e}{n_{oe}}\right)$ approximately follows. Thus the vortical motion begins, as soon as the electron density perturbation, $\delta n_e$, appears.

We use that, as it will be shown below, the characteristic frequencies of perturbations approximately equal to ion plasma frequency, $\omega_{pi}$.

As beam ions have large mass and propagate through system with velocity $V_{bi}$, we describe their dynamics in linear approximation. We derive ion density perturbation from eq.s (2)

$$\delta n_i = -n_{io}\left(\frac{q_i}{m_i}\right)\frac{\Delta \phi}{(\omega - k_z V_{ib} + i\nu_i)^2} \quad (8)$$

Here $k$, $\omega$ are wave number and frequency of perturbation, $V_{bi}$ is the unperturbed longitudinal velocity of the ions. Substituting (8) in Poisson equation (3), one can obtain

$$\frac{\beta \Delta \phi}{4\pi e} = \delta n_e, \quad \beta = 1 - \frac{\omega_{pi}^2}{(\omega - k_z V_{ib} + i\nu_i)^2},$$
$$n_e = n_{oe} + \delta n_e. \quad (9)$$

Let us consider instability development in linear approximation. Then we search the dependence of the perturbation on $z$, $\theta$ in the form $\delta n_e \propto \exp(ik_z z + i\ell_\theta \theta)$. Then from (4) we derive

$$d_t\left(\frac{\omega_{He}}{n_e}\right) = \alpha \frac{\nu_e}{n_e} - \left(\frac{e\omega_{He}}{m_e n_{oe}}\right)\frac{ik_z^2 \phi}{(\omega - \ell_\theta \omega_{\theta o} + i\nu_e)},$$
$$\omega_{\theta o} \equiv \frac{V_{\theta o}}{r}. \quad (10)$$

From (5), (6), (9), (10) we obtain, using the radial gradient of the short coil magnetic field, the following linear dispersion relation, describing the instability development

$$1 - \frac{\omega_{pi}^2}{(\omega - k_z V_{bi} + i/\tau_i)^2} - \frac{(\ell_\theta/r)\partial_r(\omega_{pe}^2/\omega_{He})}{(\omega - \ell_\theta \omega_{\theta o} + i/\tau_e)k^2} -$$
$$- \frac{\omega_{pe}^2}{(\omega - \ell_\theta \omega_{\theta o} + i/\tau_e)^2}\frac{k_z^2}{k^2} = 0 \quad (11)$$

From (11) for quick $V_{ph} \approx V_{\theta o}$ vortical perturbations we obtain

$k_z = 0$, $\omega = \omega^{(o)} + \delta \omega$, $|\delta\omega| \ll \omega^{(o)}$

$$\omega^{(o)} = \omega_{pi} = \ell_\theta \omega_{\theta o}, \quad \omega_{\theta o} = \left(\frac{\omega_{pe}^2}{2\omega_{He}}\right)\left(\frac{\Delta n}{n_{oe}}\right),$$

$$\Delta n \equiv n_{oe} - \frac{q_i n_{oi}}{e}, \quad \delta\omega = i\gamma_q,$$

$$\gamma_q \approx \left(\frac{1}{k}\right)\sqrt{\left(\frac{\omega_{pi}}{2}\right)\left(\frac{\ell_\theta}{r}\right)\left|\partial_r\left(\frac{\omega_{pe}^2}{\omega_{He}}\right)\right|} - \frac{1}{2}\left(\frac{1}{\tau_e} + \frac{1}{\tau_i}\right). \quad (12)$$

From (11) for slow $V_{ph} \ll V_{\theta o}$ vortical perturbations we obtain

$$\gamma_s = \gamma_{s0} - \frac{1}{3}\left(\frac{1}{\tau_e} + \frac{2}{\tau_i}\right), \quad \gamma_{s0} \approx \left(\frac{\sqrt{3}}{2^{4/3}}\right)[\omega_{pi}^2 \ell_\theta \omega_{\theta o}]^{1/3}$$

$$k^2 = -\left(\frac{1}{V_{\theta o}}\right)\partial_r\left(\frac{\omega_{pe}^2}{\omega_{He}}\right), \quad \text{Re}\,\omega_s = \frac{\gamma_{s0}}{\sqrt{3}}. \quad (13)$$

One can see that $\tau_e$ and $\tau_i$ decrease growth rates and lead to appearance of thresholds of instability development.

Let us consider now, how finite $k_z \neq 0$ influences on growth rate of the instability development. From (11) we obtain the growth rate of the excitation of slow homogeneous turbulence with taking into account $k_z$

$$\gamma_s \approx \left(\frac{\sqrt{3}}{2^{4/3}}\right)\omega_{pi}^{2/3}(\ell_\theta \omega_{\theta o} - k_z V_{bi})^{1/3} \times$$
$$\times \left\{1 - \frac{k_z^2}{\left[2k_z^2 + \left(\frac{\ell_\theta}{r}\right)(\ell_\theta \omega_{\theta o} - k_z V_{bi})\left|\partial_r\left(\frac{1}{\omega_{He}}\right)\right|\right]}\right\}^{1/3}. \quad (14)$$

From (14) one can see that both the taking into account the longitudinal dynamics of ions and electrons results in reduction of the growth rate. The perturbations with least $k_z \left(\approx \frac{\pi}{L}\right)$ have maximum growth rate, that is with the largest longitudinal dimensions, close to system length.

### 3. CONCLUSION

The dispersion equation, describing the instability development of vortex turbulence excitation in cylindrical plasma in crossed radial electric and axial magnetic fields with taking into account the longitudinal inhomogeneity and finite time of leaving of plasma electrons and ions from the system, has been derived. It is shown that the finite length of system time and finite time of system leaving of plasma electrons and ions leads to the appearance of the instability threshold and to decrease of growth rate of its development.